# VALIDITY AND RELIABILITY OF FREE SOFTWARE FOR BIDIMENTIONAL GAIT ANALYSIS

Ana Paula Quixadá, PT, Msc[a]; Andrea Naomi Onodera, PT, Msc[b]; Norberto Peña, PT, PhD[c]; José Garcia Vivas Miranda, PhD[d]; Katia Nunes Sá, PT, PhD[e].

a Department of Postgraduate, Master in Technology in Health, Escola Bahiana de Medicina e Saúde Pública, Salvador, Bahia, Brazil.

b Laboratory of Biomechanics, Dass Nordeste Calçados e Artigos Esportivos S.A, Ivoti, Rio Grande do Sul, Brazil.

c Department of Physical Therapy, Federal Universisty of Bahia, Salvador, Bahia, Brazil.

d Department of Physics, Federal Universisty of Bahia, Salvador, Bahia, Brazil.

e Department of Postgraduate, Master in Technology in Health, Escola Bahiana de Medicina e Saúde Pública, Salvador, Bahia, Brazil.

Corresponding Author: Ana Paula Quixadá, Rua Manoel Barreto, 354, apt. 601, Graça, Salvador, Bahia, Brazil, Phone: 07199331135, e-mail: apq.fisio@gmail.com

Acknowledgments: We want to thank the Bahia State Research Support Foundation for the academic financial support, Dass Nordeste Calçados e Artigos Esportivos SA for allowing the use of their facilities and all members of NITRE (UFBA) and Neuromuscoloskeletic Dynamics (EBMSP) research groups.

Abstract

Despite the evaluation systems of human movement that have been advancing in recent decades, their use are not feasible for clinical practice because it has a high cost and scarcity of trained operators to interpret their results. An ideal videogrammetry system should be easy to use, low cost, with minimal equipment, and fast realization. The CvMob is a free tool for dynamic evaluation of human movements that express measurements in figures, tables, and graphics. This paper aims to determine if CvMob is a reliable tool for the evaluation of two-dimensional human gait. This is a validity and reliability study. The sample was composed of 56 healthy individuals who walked on a 9-meter-long walkway and were simultaneously filmed by CvMob and Vicon system cameras. Linear trajectories and angular measurements were compared to validate the CvMob system, and inter and intrarater findings of the same measurements were used to determine reliability. A strong correlation ($r_s$ mean = 0.988) of the linear trajectories between systems and inter and intrarater analysis were found. According to the Bland-Altman method, the angles that had good agreement between systems were maximum flexion and extension (stance and swing) of the knee and dorsiflexion range of motion and stride length. The CvMob is a reliable tool for analysis of linear motion and lengths in two-dimensional evaluations of human gait. The angular measurements demonstrate high agreement for the knee joint; however, the hip and ankle measurements were limited by differences between systems.

Key words: Validity; Gait; Evalutation; Videogrammetry; CvMob.

Introduction

Human locomotion is a functional task that requires complex interactions and coordination of the nervous and musculoskeletal systems mainly involving the lower limbs, and it is one of the most important functions of the body[1].

The gait parameters are widely used in research as a primary outcome in studies on efficacy, safety, and quality of different intervention[2,3]. In clinical practice, gait evaluation is important to monitor the development of disorders[4] and the responses to implemented therapeutic modalities[2,3,5]. This assessment is essential for functional diagnosis and monitoring individuals with orthopedic, rheumatological, and neurological problems[6] and also serve as a basis for making prostheses and building robotic exoskeletons[7,8,9].

Despite its importance, the evaluation of locomotion in clinical practice is still commonly performed by the rater observation, which has a very subjective character as it is experience dependent and prone to error, thus leading to low or moderate reliability[10,11,12]. In past decades, computational systems were created to quantitatively evaluate the human gait by measuring kinetic and kinematic parameters[13,14,15] in different populations both for clinical application and to improve performance[5,2,16]. The three-dimensional analysis is an important tool for measuring human movement because it evaluates movement in all three movement planes and is a very reliable source of measurement; thus, it is considered the gold standard in many research studies[17,18,19,20]. However, the high cost and lack of human resources able to operate them and interpret their results makes it difficult to implement this technology in hospitals and outpatient clinical practices[21,6,15].

The bi-dimensional analysis is limited because it measures the movement in only one plane. However, the uniplanar analysis is a practical, simple, and inexpensive alternative to gait analysis, which is crucial to the dissemination of the quantitative gait analysis in clinical practice[6,15,22].

The CvMob is a free assessment tool of dynamic movement that expresses the results of these measurements in numbers, tables, and graphics[23], and this may be a more accurate and sensitive assessment tool than subjective evaluation performed by the rater observation. Accomplishing a validation process is necessary to guarantee reliable measurements and establish the tool for wide use. That is why the aim of this paper was to verify if the CvMob is a reliable tool for a two-dimensional analysis of human gait.

Methods

*Study Design and Participants*

This is a validation study and the sample calculation was based on 10 individuals per item evaluated[24]. Sixty healthy subjects of both genders, aged between 20 and 59 years old, were included. People who reported sensory, pain, or balance disorders at the time of assessment were excluded. The study was approved by the Ethics Committee of the Bahia School of Medicine and Public Health (CAAE: 13429113.6.0000.5544) and only the subjects who signed the consent form were part of the research.

*Procedure*

The gait evaluations were made at the Laboratory of Biomechanics of a shoe manufacturer company, Dass Nordeste Calçados e Artigos Esportivos S.A., in the state of Rio Grande do Sul in the south region in Brazil. In this laboratory, the volunteers were instructed to walk on a 9-meter-long and 1.70-meter-wide walkway. The subjects were instructed to attend the study wearing top, shorts, or swim suits.

The protocol of the lower limbs marker placement of the biomechanical model, Plug-in Gait, half body, was used, and the markers were Vicon´s 14 mm diameter reflective markers. The right greater trochanter, lateral epicondyle of the right femur, right lateral malleolus, and head of the right fifth metatarsal were marked for the CvMob analysis. The volunteer was instructed to walk barefoot on the track at a comfortable and usual speed. First for three consecutive times to ensure habituation to the environment and to the ground and then five more times for simultaneously recording videos in both instruments.

*Outcome Measures*

The camera used to record the videos used in CvMob system was the GoPro HERO 3 black edition (San Mateo, California, USA) set to narrow mode with a 1280/720 pixels (720p) resolution and 120 frames/sec. The GoPro camera was attached to a tripod and positioned at a distance of 238 cm from the middle of the walkway and a height of 79.5 cm. The CvMob calibration was always made at the beginning of all videos, using the same plane and distance of volunteers in relation to the camera. The instrument used to calibrate the system was a ruler with two Styrofoam hemispheres fixed on the ruler and separated by a distance of 20 centimeters between them.

The three-dimensional motion capture system was composed of six infrared cameras (Model T40, Vicon Motion Systems Ltd., UK) that were fixed in the laboratory ceiling and operating at a frequency of 240Hz. The filter used in Nexus software was the Butterworth, with a cut-off frequency of 6Hz and 4th order filter.

The variables were the maximum hip extension and flexion angles, maximum flexion and extension of the knee, range of motion (ROM) of dorsiflexion and plantar flexion, the stride length, and the linear trajectories of the right knee and right ankle.

The two-dimensional systems analysis is limited by the loss of points, which happens when the anatomical point is covered. This situation occurred when the upper limb, in balance, covered the hip marker; thus, it was necessary make adjustments in hip and knee angles.

CvMob measured the absolute hip angles through a coordinate system introduced by the program as a vertical reference. This is a static reference and can only be performed on one frame at a time; thus, the tracking of anatomical points by the program became impossible. Consequently we chose the frame of the maximum range of hip flexion and extension. The maximum flexion was measured on the 3rd frame prior to initial contact of the right heel and the maximum extension of the left member's initial contact.

Once the right upper limb, in balance, covered the greater trochanter marker, tracking the knee's mid-stance maximum extension and initial swing maximum flexion began in the mid-stance phase, immediately after the hand covered the marker, and ended when the upper limb returned and covered the point of the hip. The final tracking of the swing phase maximum extension started when the points of the knee angle were selected, immediately after the hand of the individual passed the anatomical hip marker.

The ankle range of motion (ROM) calculation was made from the subtraction of the first maximum plantar flexion with the maximum dorsiflexion in support and the result was the ROM of dorsiflexion. To set the plantiflexion ROM, the maximum dorsiflexion in stance phase was subtracted from the maximum value of the subsequent plantar flexion and then defined by the plantar ROM. The stride length was measured by the distance between the first and second initial contact of the right calcaneus marker.

*Data Processing*

To compare trajectories in both systems, the output data was rescaled and re-referenced. An R-script program was done to automatize the process. The angular data generated by Nexus® (Vicon Motion Systems Ltd., UK) started and stopped at the mid-swings before and after the main gait cycle. These data were processed by Origin 9.0® software (Northampton, Massachusetts, USA) that plotted graphics to identify the angles. The Nexus angle's data was subsequently placed in the database, along with the angles obtained from CvMob. The stride length data was generated by Polygon® system (Vicon Motion Systems Ltd. UK).

*Inter and Intrarater Reliability*

Two evaluators were selected. One was a physical therapist researcher with 7 years of experience in photogrammetry and the other is a physical therapist with 15 years of clinical experience in observational gait analysis. The last video of all individuals was assessed for interrater analysis, and the last video of the last 10 research subjects was selected for intrarater assessment. Each rater performed two analyses with a 7-day interval between them. Both raters were trained to use the software by a member of the CvMob® developer group. The raters' analyses were performed independently and without the knowledge of the gold standard method results.

*Statistical Analysis*

Data were tabulated using the Statistical Package for Social Sciences (SPSS) version 14.0 for Windows. The Bland-Altman method25 was used to test the angle validity between the instruments, and inter intrarater reliability. Four criteria were established to consider the agreement between variables: (1) The points should remain within the agreement interval, (2) The average of the differences value should be close to zero, (3) Agreement interval should be around the bias, and this would also vary, depending on the analyzed joint, and (4) Distribution of individuals should be close to the zero and be biased and away from the limits of agreement.

The intraclass correlation coefficient[26] was also used for intra and interrater angles analyses, which in this study has the following interpretation: less than 0.20 was considered poor agreement, 0.21 to 0.40 was acceptable, 0.41 to 0.60 was moderate, 0.61 to 0.80 was good and > 0.80 was very good[10]. The interpretation of the Spearman correlation coefficients were as follows: $r_s$ < .4 (weak correlation magnitude), $r_s$ > .4 to r < .5 (moderate magnitude), and $r_s$ >.5 (strong magnitude)[27]. A 5% margin of error and 95% confidence interval were considered.

Results

The 75 individual gaits were recorded between January 2015 and February 2015. One of the subjects was excluded for having a sensory disorder. Eighteen subjects were excluded from the angular and linear trajectory analysis by mistake between the two measurement systems in the data taken from the Nexus® and loss of points in the CvMob software. According to inter and intrarater evaluations, 13 subjects were excluded in the angular analysis and 9 individuals in the analysis of the linear trajectory due to the loss of points during the evaluation. Since the comparison of methods is the primary outcome, the demographic data represent the 56 individuals (52.6% men) of the validity test's final sample. The subject's mean weight was 72.45 $\pm$ 13.95 kg, the mean age was $31.50 \pm 8.75$ years old and their average body mass index was $25.43 \pm 3.63$ kg/cm$^2$.

The CvMob demonstrated a strong correlation of the X and Y knee and ankle linear trajectories analysis between the measurement systems (Figures 1, 2, 3 and 4). The CvMob also demonstrated a strong inter ($r_s$ mean = 0.99) and intrarater ($r_s$ mean = 0.99) correlation, which shows that the linear trajectories measured by CvMob are reproducible by different people and by the same person at different times.

The angular analysis results of the two systems, performed by the Bland-Altman method, demonstrated high agreement between the knee flexion (Figure 5) and extension angles, both in balance and in mid-stance, dorsiflexion range of motion, and stride length.

A high level of agreement between the angles was also evidenced in the two analyses of the same examiner and ratings between examiners. Only the assessment of plantar flexion of Rater 2 was considered moderate agreement because, although most of the data were in concordance interval, the distribution was not close to zero and the limits of upper and lower agreement were not near bias.

The intra and inter-rater analysis performed by intraclass correlation coefficient corroborate the Bland-Altman method results. There were only two good measures (ICC = 0.781 and 0.786) in hip flexion angle and dorsiflexion range of motion, and a good measure (ICC = 0.786) was obtained by Examiner 2 for the hip extension angle, while all other variables reached values above 0.80 and 0.90, which means a very good correlation.

Discussion

The results show that the knee angle and linear trajectory of the x and y axes data measured by CvMob are consistent with the trajectory in the sagittal plane of three-dimensional system measures. Peña and collaborators[23] validated the CvMob by comparing the linear trajectory data of a pendulum movement of the system with a theoretical model, which confirmed the CvMob's measurement accuracy. Despite this validation, there was a need to test this measurement in a human model valid situation, so the software could be used with confidence in clinics and clinical research.

The stride length had a high agreement between CvMob and the three-dimensional system and high reliability in inter and intra-examiner evaluations. There are other systems that analyze the stride length, which have moderate reliability, such as GaitMat II (ICC = 0.24)[28], and high reliability, such as GAITRite (ICC = 0.99)[29,19]. These systems analyze the spatiotemporal variables, which are part of the gait parameters. Using CvMob, the evaluator can, in one video, measure angles, linear trajectories, and stride length.

In the reliability analysis of the angular measurements, the best agreement between the systems was observed in the angles of flexion and extension of the knee (swing and support), and dorsiflexion range of motion. Ugbolue et al[28] validated the two-dimensional system based on the augmented video portable system (AVPS) and showed good results in inter and intrarater analyses and also did not find any differences between the AVPS and three-dimensional analysis.

Like AVPS, the GaitGrabber[18] is a reliable system for the spatio-temporal and angular gait data in the sagittal plane, once there were no differences with the three-dimensional system analysis and the majority of ICC values were excellent (ICC > 0.84). The CvMob is simple as these two-dimensional systems, free and does not need to have a connected camera, which enables filming in different environments, such as underwater.

In the current study, both the maximum hip flexion and the maximum extent demonstrated low concordance with the three-dimensional method. The CvMob hip angles analysis were performed based on visual selection of frames in which the absolute angles were calculated. A possible explanation for the lack of agreement in hip angles is the analysis method because the absence of the tracking by the system obligates the evaluator to arbitrarily choose the frame, which can be a source of errors. Another explanation is the limitation of the bi-dimensional analysis itself because the measurement in one plane may not be very precise due to the lack of information about the rotation movements in other planes[30]. However, when clinicians use the analysis to compare results before and after an intervention, the limitation is the same in both results, so this limitation is not important in this situation.

The plantar flexion range of motion had a low agreement between systems. In Hu-m-na system[17] there was also a lack of reliability in the ankle angle, which was justified by the marker placement error. In the GaitGrabber validation study, the ankle angle was the most reliable and the ROMs were defined from the subtraction of movement relative to the neutral position. The anatomic points that composed the ankle angle were the head of the fibula, lateral malleolus, and the head of the fifth metatarsal. Thus, one possible explanation for the lack of agreement may be the marker placement because in the pre-swing phase, the knee flexion may have interfered with the plantar flexion angle. Another possibility is the influence of camera lens deformation. Despite the actions taken to improve the image distortion, the image's peripheral areas still have distortion exactly where the subject's foot is displayed. In plantarflexion movement, the moving point is the foot, so image distortion can interfere in numerical results.

Despite the low agreement among hip angles and plantar flexion, it is important to note that the high reliability in intrarater analysis shows that if the clinician performs the test before or after an intervention is possible to quantify reliably if the treatment was successful or not. Additionally, the clinician can monitor the progress in the presence or absence of an intervention, reporting the worsening of the patient's motor condition and from this variability indicate the severity of disease[4]. The gait assessment should have a good cost-effectiveness ratio, which means that the benefit to testing must be greater than the costs for its realization. Although the three-dimensional gait analysis is the gold standard for the motion analysis, sometimes it is not economically efficient because the instruments are expensive, the time for the procedure is long, the results are not easy to interpret by the clinician, and the analysis time is long[8]. Thus it is necessary to have a system that is simple to use and low cost but also has a good reliability in generating results.

The CvMob proved to be reliable and valid for kinematic analysis in the sagittal plane of gait. It can be an accessible option and the assessment of human movement in clinical practice could become more objective, reducing the chance of errors during the evaluation and making it independent of the examiner's experience. Future studies could be done to evaluate the validity and reliability of this software in other important planes of gait, like the frontal plane. Disorders in this plane are important to monitor for some orthopedic pathologies.

Conclusion

From the results of this study, we conclude that the CvMob is a reliable tool for linear motion analysis and spatial measurements, once the measurements had high agreement and strong correlations with the three-dimensional analysis. It is also reliable for angular analysis of the knee, but for the hip and ankle angles caution is needed with the method of analysis as well as the marking of anatomical points, as these can interfere with the final result. The results generated by CvMob can be reproduced more than once, and the system can be used by several evaluators.

Conflict of Interest statement

All authors declare that there is no financial or personal interest of any nature or kind in any product, service or company that could be influencing the results presented in this manuscript.

Figures

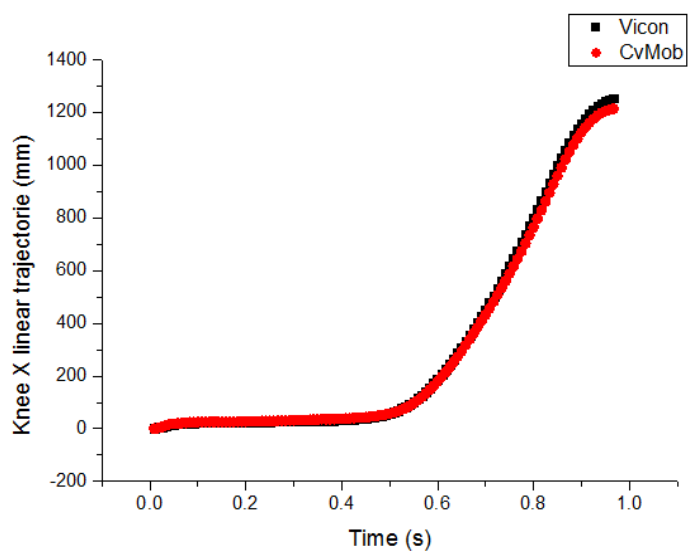

Figure 1: Example of knee X linear trajectory strongest correlation ($r_s$ = 0.999) measured by CvMob (in red) and the Vicon (in black).

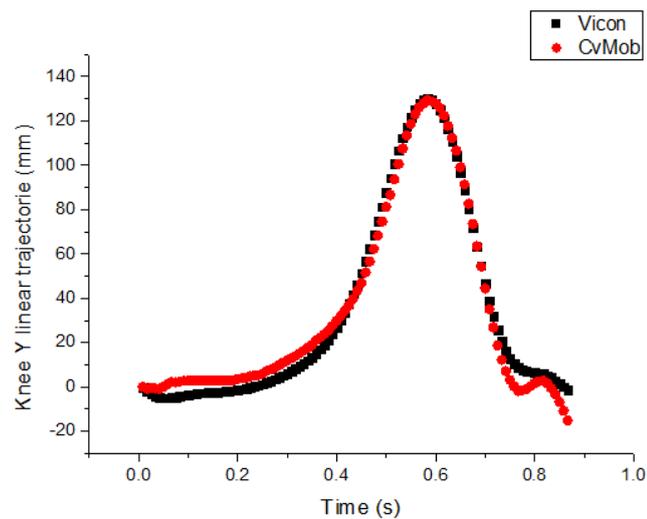

Figure 2: Example of knee Y linear trajectory strongest correlation ($r_s$ = 0.986) measured by CvMob (in red) and the Vicon (in black)

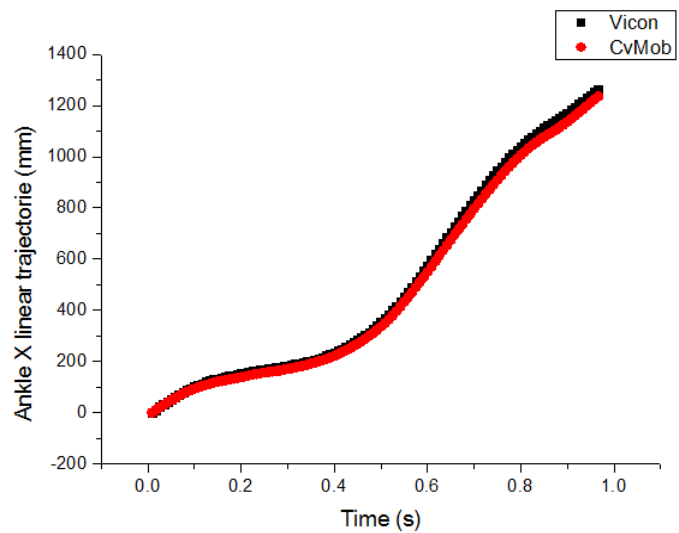

Figure 3:

Example of ankle X linear trajectory strongest correlation ($r_s = 1.00$) measured by CvMob (in red) and the Vicon (in black).

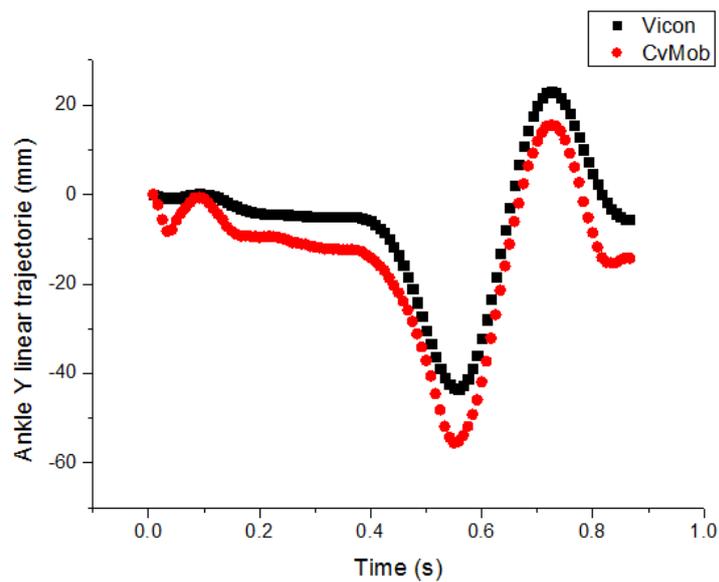

Figure 4: Example of ankle Y linear trajectory strongest correlation ($r_s = 0.986$) measured by CvMob (in red) and the Vicon (in black).

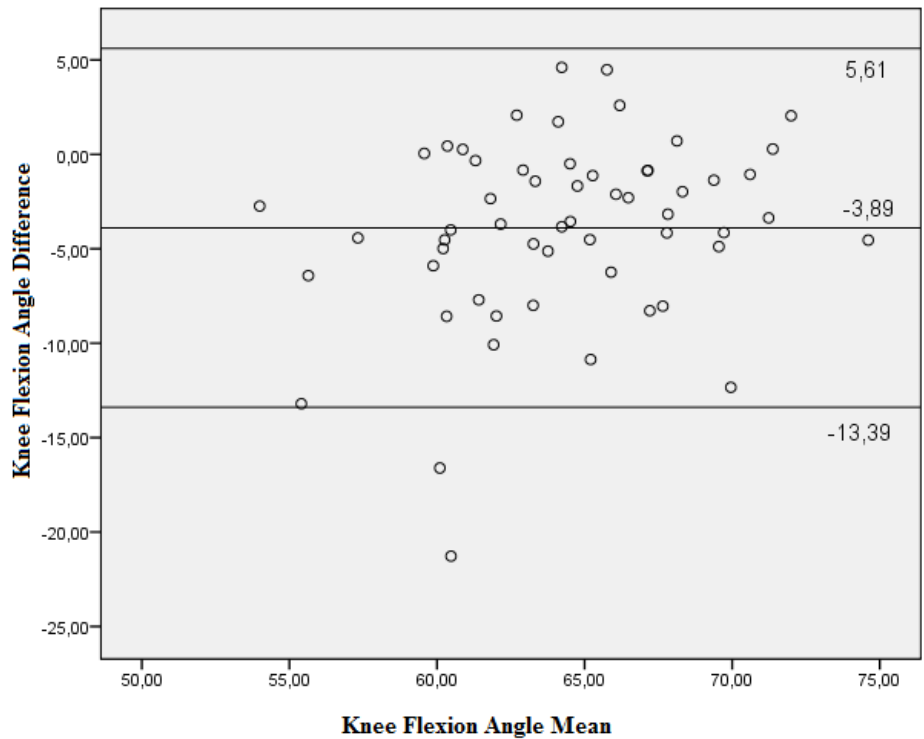

Figure 5: Scatter plot for the difference and average maximum knee flexion angle between the CvMob and Nexus;

Table 1: Intra and interrater analysis of angles and stride lenght with the intraclass correlation coeficiente.

ICC: Intraclass Correlation Coefficient; CI: Confidence Interval; ROM: Range of motion

| Angle | Intrarater Analysis: Rater 1 | | Intrarater Analysis: Rater 2 | | Interrater Analysis | |
|---|---|---|---|---|---|---|
| | ICC | IC$_{95\%}$ | ICC | IC$_{95\%}$ | ICC | IC$_{95\%}$ |
| Hip flexion | .997 | .986 − .999 | .879 | .504 − .970 | .781 | .636 − .868 |
| Hip extension | .864 | .427 − .966 | .786 | .112 − .947 | .822 | .705 − .893 |
| Knee extension in midstance | .996 | .984 − .999 | .961 | .851 − .990 | .974 | .956 − .984 |
| Knee flexion | .992 | .968 − .998 | .856 | .462 − .964 | .931 | .885 − .958 |
| Knee extension in final swing | .985 | .943 − .996 | .966 | .865 − .991 | .948 | .914 − .969 |
| Dorsiflexion ROM | .989 | .956 − .997 | .891 | .539 − .973 | .786 | .604 − .880 |
| Plantar flexion ROM | .977 | .899 − .994 | .936 | .737 − .984 | .881 | .771 − .934 |
| Stride lenght | .998 | .994 − 1.00 | .982 | .929 − .996 | .991 | .984 − .994 |

Table 2: Correlation of X and Y linear trajectories of the knee and ankle between the Nexus® system and the CvMob®.

| Subjects | Knee X | | Knee Y | | Ankle X | | Ankle Y | |
|---|---|---|---|---|---|---|---|---|
| | $\rho$ | $p*$ | $\rho$ | $p*$ | $\rho$ | $p*$ | $\rho$ | $p*$ |
| S001 | 0,999 | <0,001 | 0,882 | <0,001 | 1,000 | <0,001 | 0,694 | <0,001 |
| S002 | 0,997 | <0,001 | 0,897 | <0,001 | 1,000 | <0,001 | 0,883 | <0,001 |
| S003 | 0,997 | <0,001 | 0,745 | <0,001 | 1,000 | <0,001 | 0,908 | <0,001 |
| S004 | 0,996 | <0,001 | 0,945 | <0,001 | 1,000 | <0,001 | 0,964 | <0,001 |
| S005 | 0,999 | <0,001 | 0,815 | <0,001 | 1,000 | <0,001 | 0,902 | <0,001 |
| S006 | 0,999 | <0,001 | 0,933 | <0,001 | 1,000 | <0,001 | 0,920 | <0,001 |
| S007 | 0,999 | <0,001 | 0,965 | <0,001 | 1,000 | <0,001 | 0,922 | <0,001 |
| S011 | 0,999 | <0,001 | 0,926 | <0,001 | 1,000 | <0,001 | 0,964 | <0,001 |
| S012 | 0,998 | <0,001 | 0,833 | <0,001 | 1,000 | <0,001 | 0,936 | <0,001 |
| S017 | 0,999 | <0,001 | 0,883 | <0,001 | 1,000 | <0,001 | 0,957 | <0,001 |
| S018 | 0,998 | <0,001 | 0,986 | <0,001 | 1,000 | <0,001 | 0,970 | <0,001 |
| S020 | 0,998 | <0,001 | 0,842 | <0,001 | 1,000 | <0,001 | 0,912 | <0,001 |
| S021 | 0,995 | <0,001 | 0,810 | <0,001 | 1,000 | <0,001 | 0,931 | <0,001 |
| S022 | 0,998 | <0,001 | 0,842 | <0,001 | 1,000 | <0,001 | 0,917 | <0,001 |
| S024 | 0,999 | <0,001 | 0,952 | <0,001 | 1,000 | <0,001 | 0,931 | <0,001 |
| S025 | 0,999 | <0,001 | 0,971 | <0,001 | 1,000 | <0,001 | 0,936 | <0,001 |
| S026 | 0,998 | <0,001 | 0,824 | <0,001 | 1,000 | <0,001 | 0,897 | <0,001 |

| Subjects | | | | | | | | |
|---|---|---|---|---|---|---|---|---|
| S027 | 0,993 | <0,001 | 0,815 | <0,001 | 1,000 | <0,001 | 0,854 | <0,001 |
| S028 | 0,995 | <0,001 | 0,959 | <0,001 | 1,000 | <0,001 | 0,929 | <0,001 |
| S029 | 0,999 | <0,001 | 0,904 | <0,001 | 1,000 | <0,001 | 0,985 | <0,001 |
| S030 | 0,999 | <0,001 | 0,868 | <0,001 | 1,000 | <0,001 | 0,961 | <0,001 |
| S032 | 0,999 | <0,001 | 0,883 | <0,001 | 1,000 | <0,001 | 0,943 | <0,001 |
| S033 | 0,996 | <0,001 | 0,885 | <0,001 | 1,000 | <0,001 | 0,943 | <0,001 |
| S034 | 0,999 | <0,001 | 0,959 | <0,001 | 1,000 | <0,001 | 0,969 | <0,001 |
| S035 | 0,999 | <0,001 | 0,798 | <0,001 | 1,000 | <0,001 | 0,937 | <0,001 |
| S036 | 0,996 | <0,001 | 0,845 | <0,001 | 1,000 | <0,001 | 0,924 | <0,001 |
| S038 | 0,992 | <0,001 | 0,802 | <0,001 | 1,000 | <0,001 | 0,853 | <0,001 |
| S039 | 0,999 | <0,001 | 0,831 | <0,001 | 1,000 | <0,001 | 0,95 | <0,001 |
| S040 | 0,999 | <0,001 | 0,791 | <0,001 | 1,000 | <0,001 | 0,836 | <0,001 |
| S041 | 0,999 | <0,001 | 0,856 | <0,001 | 1,000 | <0,001 | 0,698 | <0,001 |
| S044 | 0,998 | <0,001 | 0,791 | <0,001 | 1,000 | <0,001 | 0,814 | <0,001 |
| S045 | 0,982 | <0,001 | 0,816 | <0,001 | 1,000 | <0,001 | 0,876 | <0,001 |
| S046 | 0,997 | <0,001 | 0,796 | <0,001 | 1,000 | <0,001 | 0,845 | <0,001 |
| S047 | 0,998 | <0,001 | 0,868 | <0,001 | 1,000 | <0,001 | 0,731 | <0,001 |
| S048 | 0,999 | <0,001 | 0,713 | <0,001 | 1,000 | <0,001 | 0,869 | <0,001 |
| S050 | 0,998 | <0,001 | 0,769 | <0,001 | 1,000 | <0,001 | 0,728 | <0,001 |
| S051 | 0,998 | <0,001 | 0,876 | <0,001 | 1,000 | <0,001 | 0,749 | <0,001 |
| S052 | 0,997 | <0,001 | 0,856 | <0,001 | 1,000 | <0,001 | 0,662 | <0,001 |
| S054 | 0,997 | <0,001 | 0,785 | <0,001 | 1,000 | <0,001 | 0,889 | <0,001 |
| S055 | 0,997 | <0,001 | 0,834 | <0,001 | 1,000 | <0,001 | 0,681 | <0,001 |
| S056 | 0,968 | <0,001 | 0,691 | <0,001 | 1,000 | <0,001 | 0,713 | <0,001 |
| S057 | 0,998 | <0,001 | 0,797 | <0,001 | 1,000 | <0,001 | 0,791 | <0,001 |
| S058 | 0,998 | <0,001 | 0,799 | <0,001 | 1,000 | <0,001 | 0,713 | <0,001 |
| S060 | 0,999 | <0,001 | 0,734 | <0,001 | 1,000 | <0,001 | 0,662 | <0,001 |
| S061 | 0,999 | <0,001 | 0,860 | <0,001 | 1,000 | <0,001 | 0,697 | <0,001 |
| S062 | 0,998 | <0,001 | 0,827 | <0,001 | 1,000 | <0,001 | 0,794 | <0,001 |
| S063 | 0,992 | <0,001 | 0,867 | <0,001 | 1,000 | <0,001 | 0,931 | <0,001 |
| S064 | 0,999 | <0,001 | 0,839 | <0,001 | 1,000 | <0,001 | 0,870 | <0,001 |
| S066 | 0,999 | <0,001 | 0,843 | <0,001 | 1,000 | <0,001 | 0,682 | <0,001 |
| S068 | 0,999 | <0,001 | 0,821 | <0,001 | 1,000 | <0,001 | 0,901 | <0,001 |
| S070 | 0,999 | <0,001 | 0,766 | <0,001 | 1,000 | <0,001 | 0,801 | <0,001 |
| S071 | 0,974 | <0,001 | 0,846 | <0,001 | 1,000 | <0,001 | 0,755 | <0,001 |
| S072 | 0,999 | <0,001 | 0,789 | <0,001 | 1,000 | <0,001 | 0,865 | <0,001 |
| S073 | 0,999 | <0,001 | 0,842 | <0,001 | 1,000 | <0,001 | 0,826 | <0,001 |
| S075 | 0,998 | <0,001 | 0,794 | <0,001 | 1,000 | <0,001 | 0,916 | <0,001 |
| S076 | 0,998 | <0,001 | 0,830 | <0,001 | 1,000 | <0,001 | 0,788 | <0,001 |

Table 3: Interrater correlation of X and Y trajectories of knee and ankle between the rater 1 and rater 2 through CvMob®.

| Subjects | Knee X | | Knee Y | | Ankle X | | Ankle Y | |
|---|---|---|---|---|---|---|---|---|
| | $\rho$ | $p*$ | $\rho$ | $p*$ | $\rho$ | $p*$ | $\rho$ | $p*$ |
| S001 | 1,000 | <0,001 | 0,997 | <0,001 | 1,000 | <0,001 | 1,000 | <0,001 |
| S002 | 1,000 | <0,001 | 0,999 | <0,001 | 1,000 | <0,001 | 1,000 | <0,001 |

| | | | | | | | |
|---|---|---|---|---|---|---|---|
| S003 | 1,000 | <0,001 | 1,000 | <0,001 | 1,000 | <0,001 | 0,991 | <0,001 |
| S004 | 1,000 | <0,001 | 1,000 | <0,001 | 1,000 | <0,001 | 0,998 | <0,001 |
| S005 | 1,000 | <0,001 | 1,000 | <0,001 | 1,000 | <0,001 | 1,000 | <0,001 |
| S006 | 1,000 | <0,001 | 1,000 | <0,001 | 1,000 | <0,001 | 1,000 | <0,001 |
| S007 | 1,000 | <0,001 | 0,998 | <0,001 | 1,000 | <0,001 | 1,000 | <0,001 |
| S009 | 1,000 | <0,001 | 1,000 | <0,001 | 1,000 | <0,001 | 1,000 | <0,001 |
| S010 | 1,000 | <0,001 | 0,996 | <0,001 | 1,000 | <0,001 | 1,000 | <0,001 |
| S011 | 1,000 | <0,001 | 1,000 | <0,001 | 1,000 | <0,001 | 1,000 | <0,001 |
| S012 | 1,000 | <0,001 | 1,000 | <0,001 | 1,000 | <0,001 | 0,998 | <0,001 |
| S013 | 1,000 | <0,001 | 0,999 | <0,001 | 1,000 | <0,001 | 0,914 | <0,001 |
| S014 | 1,000 | <0,001 | 1,000 | <0,001 | 1,000 | <0,001 | 0,999 | <0,001 |
| S015 | 1,000 | <0,001 | 0,999 | <0,001 | 1,000 | <0,001 | 1,000 | <0,001 |
| S017 | 1,000 | <0,001 | 0,998 | <0,001 | 1,000 | <0,001 | 0,997 | <0,001 |
| S018 | 1,000 | <0,001 | 1,000 | <0,001 | 1,000 | <0,001 | 0,965 | <0,001 |
| S020 | 1,000 | <0,001 | 0,983 | <0,001 | 0,997 | <0,001 | 0,991 | <0,001 |
| S021 | 1,000 | <0,001 | 1,000 | <0,001 | 1,000 | <0,001 | 1,000 | <0,001 |
| S022 | 1,000 | <0,001 | 0,993 | <0,001 | 0,993 | <0,001 | 0,863 | <0,001 |
| S024 | 1,000 | <0,001 | 1,000 | <0,001 | 1,000 | <0,001 | 0,999 | <0,001 |
| S025 | 1,000 | <0,001 | 1,000 | <0,001 | 1,000 | <0,001 | 1,000 | <0,001 |
| S026 | 1,000 | <0,001 | 1,000 | <0,001 | 1,000 | <0,001 | 0,998 | <0,001 |
| S027 | 1,000 | <0,001 | 1,000 | <0,001 | 1,000 | <0,001 | 0,999 | <0,001 |
| S028 | 1,000 | <0,001 | 1,000 | <0,001 | 1,000 | <0,001 | 1,000 | <0,001 |
| S029 | 1,000 | <0,001 | 0,999 | <0,001 | 1,000 | <0,001 | 0,993 | <0,001 |
| S030 | 1,000 | <0,001 | 1,000 | <0,001 | 1,000 | <0,001 | 0,999 | <0,001 |
| S031 | 1,000 | <0,001 | 1,000 | <0,001 | 1,000 | <0,001 | 0,985 | <0,001 |
| S032 | 1,000 | <0,001 | 1,000 | <0,001 | 1,000 | <0,001 | 1,000 | <0,001 |
| S033 | 1,000 | <0,001 | 1,000 | <0,001 | 0,999 | <0,001 | 0,871 | <0,001 |
| S034 | 1,000 | <0,001 | 0,967 | <0,001 | 0,999 | <0,001 | 0,973 | <0,001 |
| S035 | 1,000 | <0,001 | 1,000 | <0,001 | 1,000 | <0,001 | 1,000 | <0,001 |
| S036 | 1,000 | <0,001 | 1,000 | <0,001 | 0,995 | <0,001 | 0,994 | <0,001 |
| S037 | 1,000 | <0,001 | 1,000 | <0,001 | 1,000 | <0,001 | 1,000 | <0,001 |
| S038 | 1,000 | <0,001 | 0,998 | <0,001 | 1,000 | <0,001 | 1,000 | <0,001 |
| S039 | 1,000 | <0,001 | 1,000 | <0,001 | 1,000 | <0,001 | 1,000 | <0,001 |
| S040 | 1,000 | <0,001 | 1,000 | <0,001 | 1,000 | <0,001 | 1,000 | <0,001 |
| S041 | 1,000 | <0,001 | 0,981 | <0,001 | 0,999 | <0,001 | 0,960 | <0,001 |
| S044 | 1,000 | <0,001 | 1,000 | <0,001 | 1,000 | <0,001 | 1,000 | <0,001 |
| S045 | 1,000 | <0,001 | 0,991 | <0,001 | 1,000 | <0,001 | 0,999 | <0,001 |
| S046 | 1,000 | <0,001 | 0,994 | <0,001 | 0,998 | <0,001 | 0,987 | <0,001 |
| S047 | 1,000 | <0,001 | 0,995 | <0,001 | 1,000 | <0,001 | 0,999 | <0,001 |
| S048 | 1,000 | <0,001 | 0,985 | <0,001 | 1,000 | <0,001 | 1,000 | <0,001 |
| S050 | 1,000 | <0,001 | 0,993 | <0,001 | 1,000 | <0,001 | 0,990 | <0,001 |
| S052 | 1,000 | <0,001 | 0,995 | <0,001 | 0,999 | <0,001 | 0,989 | <0,001 |
| S053 | 1,000 | <0,001 | 0,950 | <0,001 | 0,993 | <0,001 | 0,975 | <0,001 |
| S054 | 1,000 | <0,001 | 1,000 | <0,001 | 1,000 | <0,001 | 1,000 | <0,001 |
| S056 | 1,000 | <0,001 | 0,988 | <0,001 | 1,000 | <0,001 | 0,998 | <0,001 |
| S057 | 1,000 | <0,001 | 0,988 | <0,001 | 0,999 | <0,001 | 0,994 | <0,001 |
| S058 | 1,000 | <0,001 | 0993 | <0,001 | 1,000 | <0,001 | 0,995 | <0,001 |
| S059 | 1,000 | <0,001 | 1,000 | <0,001 | 1,000 | <0,001 | 1,000 | <0,001 |
| S060 | 1,000 | <0,001 | 0,830 | <0,001 | 1,000 | <0,001 | 0,953 | <0,001 |
| S061 | 1,000 | <0,001 | 0,998 | <0,001 | 1,000 | <0,001 | 0,998 | <0,001 |
| S062 | 1,000 | <0,001 | 1,000 | <0,001 | 1,000 | <0,001 | 1,000 | <0,001 |
| S063 | 1,000 | <0,001 | 1,000 | <0,001 | 1,000 | <0,001 | 0,999 | <0,001 |
| S064 | 1,000 | <0,001 | 0,982 | <0,001 | 1,000 | <0,001 | 0,998 | <0,001 |
| S066 | 1,000 | <0,001 | 0,938 | <0,001 | 1,000 | <0,001 | 0,984 | <0,001 |
| S068 | 1,000 | <0,001 | 0,992 | <0,001 | 1,000 | <0,001 | 0,991 | <0,001 |
| S069 | 1,000 | <0,001 | 1,000 | <0,001 | 1,000 | <0,001 | 0,995 | <0,001 |
| S070 | 1,000 | <0,001 | 0,977 | <0,001 | 1,000 | <0,001 | 0,997 | <0,001 |
| S071 | 1,000 | <0,001 | 0,992 | <0,001 | 1,000 | <0,001 | 0,990 | <0,001 |
| S072 | 1,000 | <0,001 | 0,989 | <0,001 | 1,000 | <0,001 | 1,000 | <0,001 |
| S073 | 1,000 | <0,001 | 0,992 | <0,001 | 1,000 | <0,001 | 0,998 | <0,001 |
| S074 | 1,000 | <0,001 | 1,000 | <0,001 | 1,000 | <0,001 | 1,000 | <0,001 |
| S075 | 1,000 | <0,001 | 0,992 | <0,001 | 1,000 | <0,001 | 1,000 | <0,001 |
| S076 | 1,000 | <0,001 | 0,978 | <0,001 | 1,000 | <0,001 | 0,999 | <0,001 |

* Significance level = 5%

Table 4: Intrarater correlation of X and Y trajectories of the knee and ankle through CvMob®.

| Subjects/Joint | Rater 1 | | | | Rater 2 | | | |
|---|---|---|---|---|---|---|---|---|
| | ρX* | p*** | ρY** | p*** | ρX* | p*** | ρY** | p*** |
| S066 Knee | 1,000 | <0,001 | 0,999 | <0,001 | 1,000 | <0,001 | 0,998 | <0,001 |
| S066 Ankle | 1,000 | <0,001 | 0,998 | <0,001 | 1,000 | <0,001 | 1,000 | <0,001 |
| S068 Knee | 0,994 | <0,001 | 0,944 | <0,001 | 1,000 | <0,001 | 1,000 | <0,001 |
| S068 Ankle | 0,999 | <0,001 | 0,995 | <0,001 | 1,000 | <0,001 | 1,000 | <0,001 |
| S069 Knee | 1,000 | <0,001 | 1,000 | <0,001 | 1,000 | <0,001 | 1,000 | <0,001 |
| S069 Ankle | 1,000 | <0,001 | 1,000 | <0,001 | 1,000 | <0,001 | 1,000 | <0,001 |
| S070 Knee | 1,000 | <0,001 | 0,987 | <0,001 | 1,000 | <0,001 | 1,000 | <0,001 |
| S070 Ankle | 1,000 | <0,001 | 1,000 | <0,001 | 1,000 | <0,001 | 0,998 | <0,001 |
| S071 Knee | 1,000 | <0,001 | 0,999 | <0,001 | 1,000 | <0,001 | 0,992 | <0,001 |
| S071 Ankle | 1,000 | <0,001 | 0,999 | <0,001 | 0,999 | <0,001 | 0,982 | <0,001 |
| S072 Knee | 1,000 | <0,001 | 1,000 | <0,001 | 1,000 | <0,001 | 0,999 | <0,001 |
| S072 Ankle | 1,000 | <0,001 | 1,000 | <0,001 | 1,000 | <0,001 | 0,985 | <0,001 |
| S073 Knee | 1,000 | <0,001 | 1,000 | <0,001 | 1,000 | <0,001 | 1,000 | <0,001 |
| S073 Ankle | 1,000 | <0,001 | 1,000 | <0,001 | 1,000 | <0,001 | 0,999 | <0,001 |
| S074 Knee | 1,000 | <0,001 | 0,999 | <0,001 | 1,000 | <0,001 | 1,000 | <0,001 |
| S074 Ankle | 1,000 | <0,001 | 1,000 | <0,001 | 1,000 | <0,001 | 1,000 | <0,001 |
| S075 Knee | 1,000 | <0,001 | 1,000 | <0,001 | 1,000 | <0,001 | 1,000 | <0,001 |
| S075 Ankle | 1,000 | <0,001 | 0,998 | <0,001 | 1,000 | <0,001 | 1,000 | <0,001 |
| S076 Knee | 1,000 | <0,001 | 1,000 | <0,001 | 1,000 | <0,001 | 1,000 | <0,001 |
| S076 Ankle | 1,000 | <0,001 | 1,000 | <0,001 | 1,000 | <0,001 | 0,509 | <0,001 |

* X Spearman correlation (ρ); ** Y Spearman correlation (ρ); *** Significance level = 5%.